\documentclass[twocolumn,aps,graphicx,prl,superscriptaddress,floatfix]{revtex4-2}

\usepackage{graphicx}
\usepackage{subfigure}
\usepackage{epstopdf}
\usepackage{amsmath}
\usepackage{amssymb}
\usepackage{amsfonts}
\usepackage{amsthm}
\usepackage{mathrsfs}
\usepackage{multirow}
\usepackage{tabularx}
\usepackage{color,ulem}
\usepackage{natbib}
\usepackage[table]{xcolor}
\usepackage{booktabs}
\usepackage{colortbl}
\usepackage{tabu}
\usepackage{array}
\usepackage{ulem} 
\usepackage[colorlinks,citecolor=blue]{hyperref}

\usepackage{braket,appendix} 
\usepackage{dcolumn}
\usepackage{bm}

\begin{document}
\title{Type-II quantum spin Hall insulator}

\author{Panjun Feng}
\thanks{These authors contributed equally to this work. P. Feng performed the first-principles calculations, and C.-Y. Tan performed the model study.}
\affiliation{The Center for Advanced Quantum Studies and  School of Physics and Astronomy, Beijing Normal University, Beijing 100875, China}
\affiliation{Key Laboratory of Multiscale Spin Physics (Ministry of Education), Beijing Normal University, Beijing 100875, China}

\author{Chao-Yang Tan}
\thanks{These authors contributed equally to this work. P. Feng performed the first-principles calculations, and C.-Y. Tan performed the model study.}
\affiliation{Department of Physics and Beijing Key Laboratory of Opto-electronic Functional Materials $\&$ Micro-nano Devices. Renmin University of China, Beijing 100872, China}
\affiliation{Key Laboratory of Quantum State Construction and Manipulation (Ministry of Education), Renmin University of China, Beijing 100872, China}

\author{Miao Gao}
\affiliation{Department of Physics, School of Physical Science and Technology, Ningbo University, Zhejiang 315211, China}

\author{Xun-Wang Yan}
\affiliation{College of Physics and Engineering, Qufu Normal University, Shandong 273165, China}

\author{Zheng-Xin Liu}
\affiliation{Department of Physics and Beijing Key Laboratory of Opto-electronic Functional Materials $\&$ Micro-nano Devices. Renmin University of China, Beijing 100872, China}
\affiliation{Key Laboratory of Quantum State Construction and Manipulation (Ministry of Education), Renmin University of China, Beijing 100872, China}

\author{Peng-Jie Guo}\email{guopengjie@ruc.edu.cn}
\affiliation{Department of Physics and Beijing Key Laboratory of Opto-electronic Functional Materials $\&$ Micro-nano Devices. Renmin University of China, Beijing 100872, China}
\affiliation{Key Laboratory of Quantum State Construction and Manipulation (Ministry of Education), Renmin University of China, Beijing 100872, China}

\author{Fengjie Ma}\email{fengjie.ma@bnu.edu.cn}
\affiliation{The Center for Advanced Quantum Studies and School of Physics and Astronomy, Beijing Normal University, Beijing 100875, China}
\affiliation{Key Laboratory of Multiscale Spin Physics (Ministry of Education), Beijing Normal University, Beijing 100875, China}

\author{Zhong-Yi Lu}\email{zlu@ruc.edu.cn}
\affiliation{Department of Physics and Beijing Key Laboratory of Opto-electronic Functional Materials $\&$ Micro-nano Devices. Renmin University of China, Beijing 100872, China}
\affiliation{Key Laboratory of Quantum State Construction and Manipulation (Ministry of Education), Renmin University of China, Beijing 100872, China}
\affiliation{Hefei National Laboratory, Hefei 230088, China}

\date{\today}

\begin{abstract}

Quantum spin Hall effect is usually realized in two-dimensional materials with time-reversal symmetry, but whether it can be realized without symmetry protection remains unexplored. Here, we propose type-II quantum spin Hall insulator with quantized spin Hall conductivity, whose edge states with opposite chirality and polarization, distributed in different Brillouin zone regions, connect the conduction and valence bands at the boundary. Thus, the type-II quantum spin Hall insulator does not require any symmetry protection other than translational symmetry. Then, based on symmetry analysis and the first-principles electronic structure calculations, we demonstrate that type-II quantum spin Hall insulator can be realized in both altermagnetic materials and Luttinger compensated magnetic materials. Furthermore, based on lattice model, we find that as long as $U(1)$ symmetry exists, type-II quantum spin Hall insulator phase can always exist stably. However, if $U(1)$ symmetry is broken, type-II quantum spin Hall insulator phase transforms into an obstructed atomic insulator phase as spin-orbit coupling effect is enhanced. Therefore, our work not only proposes a new mechanism for realizing the quantum spin Hall effect, but also enriches the types of unconventional magnetic topological phases. 

\end{abstract}

\maketitle

\begin{figure}
	\begin{center}
		\includegraphics[width=1.0\columnwidth]{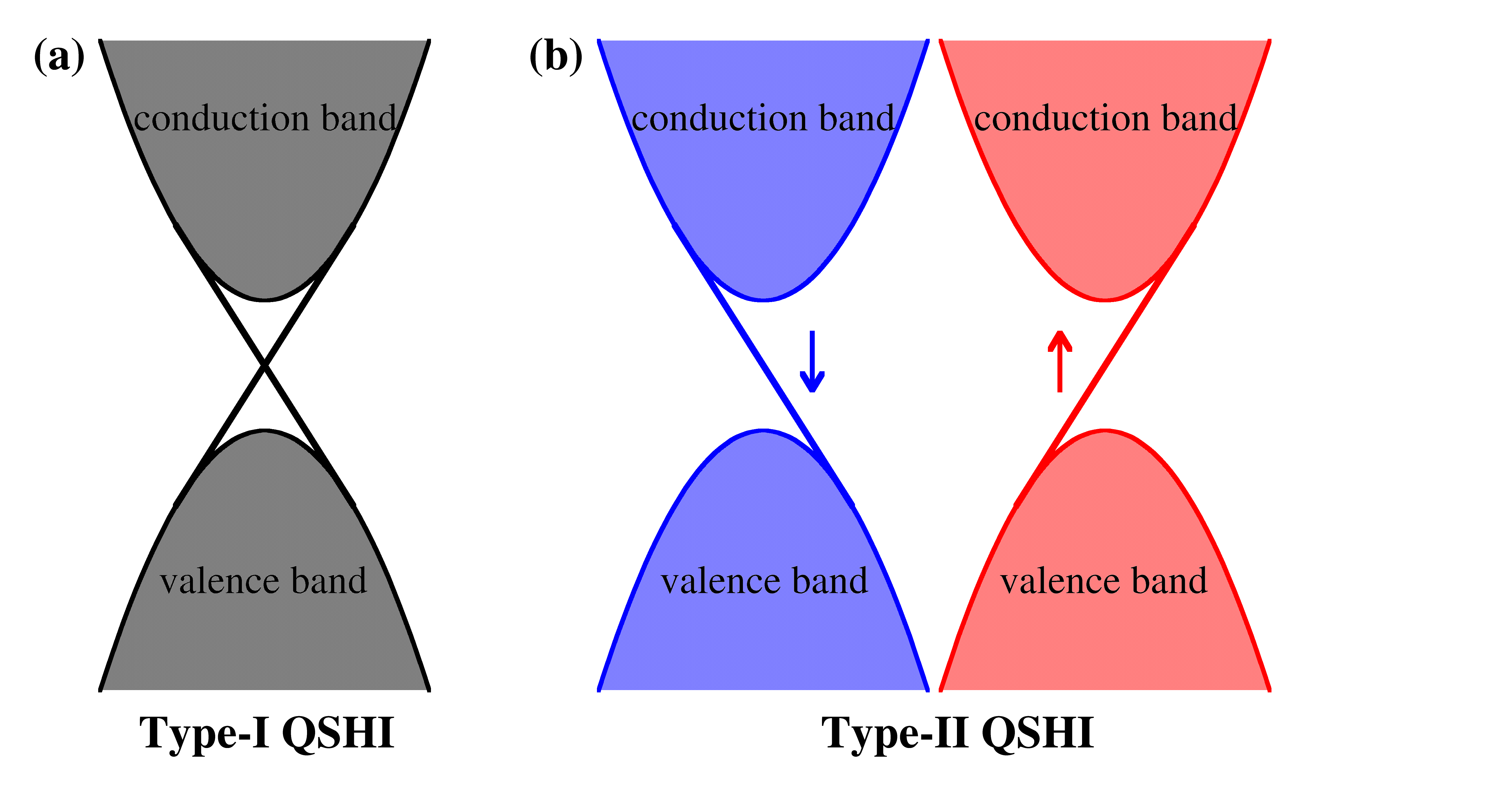}
		\caption{(a) Conventional type-I QSHI with helical edge states protected by time-reversal symmetry in non-magnetic systems. (b) Type-II QSHI with spin-polarized edge states located at different k-points in opposite spin spaces without symmetry protection.} \label{QSHI}
	\end{center}
\end{figure}

{\it Introduction.} Quantum spin Hall insulators (QSHIs) have been theoretically proposed and experimentally verified, which has sparked widespread interest in topological research \cite{PhysRevLett.95.146802, PhysRevLett.95.226801, SCZhang-QSH-theory, SCZhang-QSH-exp}. QSHIs are typically realized in two-dimensional (2D) non-magnetic materials, featuring topologically helical edge states protected by time-reversal ($\mathcal{T}$) symmetry at their boundaries as shown in Fig. \ref{QSHI}(a) \cite{RevModPhys.82.3045, RevModPhys.83.1057, RevModPhys.88.021004}. These unique properties have significant applications in future electronic devices. An open but very desire question is whether or not we can still have QSHI phase if $\mathcal{T}$ symmetry is broken? Although QSHIs have been proposed in antiferromagnets, these antiferromagnetic materials all possess equivalent $\mathcal{T}$ symmetry (either the combined symmetry of spatial inversion $\mathcal{I}$ and $\mathcal{T}$, denoted as $\mathcal{IT}$, or the combined symmetry of $\mathcal{T}$ and fractional translation $\tau$, denoted as $\mathcal{T\tau}$) \cite{PhysRevB.103.085109, PhysRevB.108.075138}. Therefore, it is not surprising that QSHI can be also realized in these antiferromagnetic materials. Thus, can QSHI phase be realized in systems where $\mathcal{T}$ symmetry is completely broken?

Recently, altermagnetism has been proposed as a novel magnetic phase distinct from ferromagnetism and antiferromagnetism \cite{altermagnetism-1, altermagnetism-2, altermagnetism-3, altermagnetism-4, PRX-1, PRX-2, PRX-3, QAH-npj2023}. Due to the fact that altermagnetic materials possess both spin splitting similar to ferromagnetic  materials and zero net magnetic moment like antiferromagnetic materials, altermagnetic materials exhibit many novel physical properties \cite{altermagnetism-1, altermagnetism-2, altermagnetism-3, altermagnetism-4, PRX-1, PRX-2, PRX-3, QAH-npj2023, PhysRevLett.134.096703, PhysRevB.110.144412,zyuzin2025,roig2024}. Based on the spin group symmetry, the anisotropic spin splittings in altermagnetic materials are categorized into three types: $d$-wave, $g$-wave, and $i$-wave \cite{PRX-2}. For $d$-wave 2D altermagnetic materials, the spins of band splitting in the $\Gamma$-X direction and $\Gamma$–Y direction can be opposite, which leads to new physical effects such as the crystal valley Hall effect and piezomagnetic effect and so on \cite{PhysRevB.111.094411}. As schematically shown by Fig. 1(b), if a $d$-wave altermagnetic material has a spin-up band inversion around the high-symmetry X point and a spin-down band inversion around the high-symmetry Y point, the spin-up and spin-down bands can contribute opposite Chern numbers under spin-orbit coupling (SOC). When the system is projected onto a one-dimensional boundary, the high-symmetry Y and X points are projected to the $\tilde{\Gamma}$ point and $\tilde{X}$ point of the one-dimensional Brillouin zone, respectively. The opposite spin Chern numbers result in opposite chirality edge states connecting the valence and conduction bands near the $\tilde{\Gamma}$ and $\tilde{X}$ points, respectively, as shown in Fig. \ref{QSHI}(b). It turns out that, the quantum spin Hall effect (QSHE)  does not require any symmetry protection other than translational symmetry. Here, we refer to this type of spin Hall insulator as a type-II QSHI, which may be realized in altermagnetic materials and even in Luttinger compensated magnetic materials, in distinction from the conventional QSHI (termed as type I).

In this study, based on symmetry analysis and the first-principles electronic structure calculations, we demonstrate that type-II QSHI without symmetry protection can be realized in both 2D altermagnetic material Nb$_2$SeTeO and Luttinger compensated magnetic material Nb$_2$SeTeO under sheer strain. Moreover, the quantized spin Hall conductivity can also be realized in Nb$_2$SeTeO without and with sheer strain. Furthermore, through lattice model calculations and analysis, we find that as long as the  $U(1)$ symmetry is present, type-II QSHI phase can be always remain stable. However, if the $U(1)$ symmetry is broken, with the enhancement of SOC, type-II QSHI phase will transition into an obstructed atomic insulator (OAI) phase \cite{OAI-1, xu2021threedimensionalrealspaceinvariants}.

\begin{figure}
	\begin{center}
		\includegraphics[width=1.0\columnwidth]{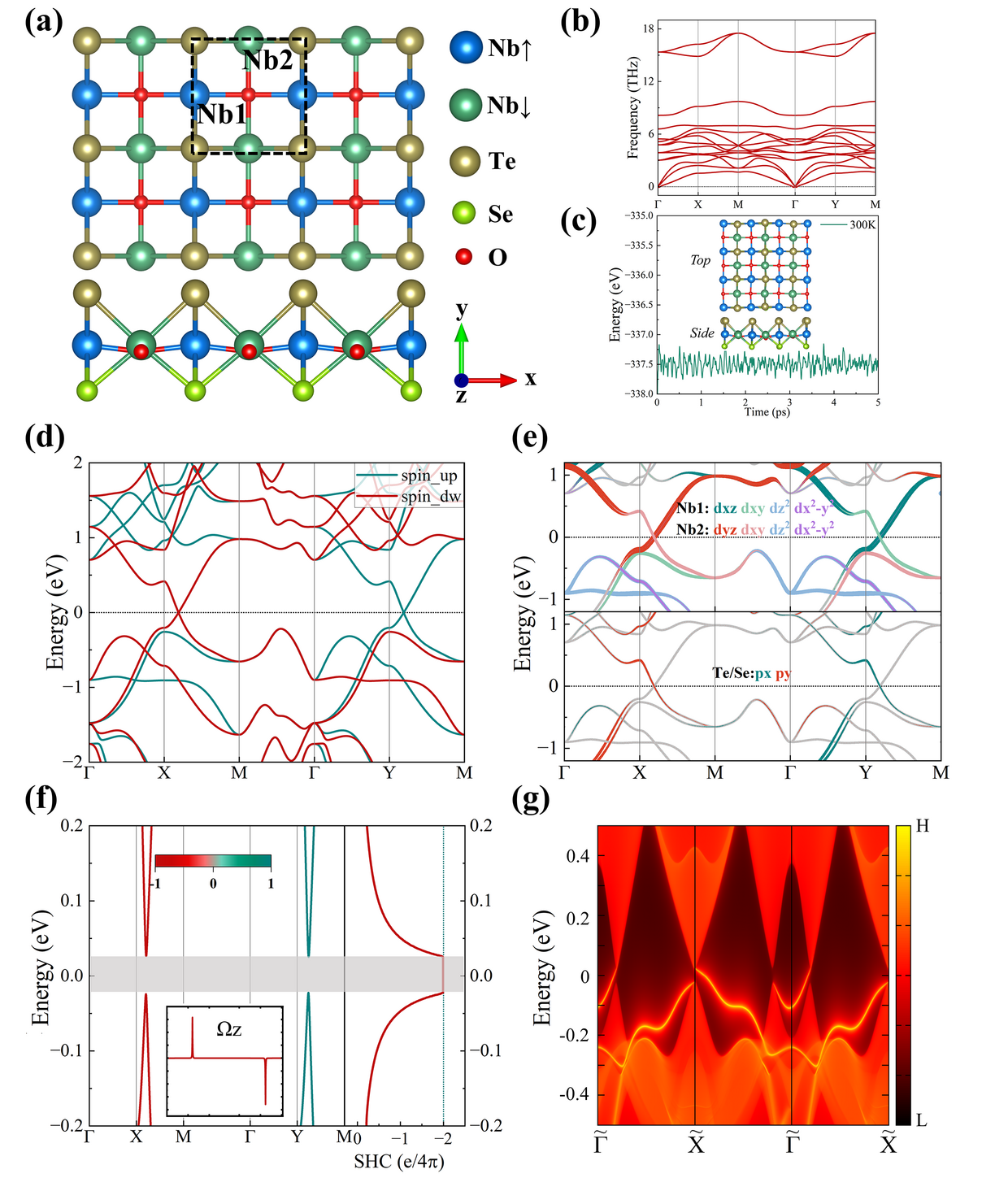}
		\caption{(a) Top view and side view of monolayer Nb$_2$SeTeO. The black dashed rectangle is the primitive cell. (b) Phonon spectra. (c) The total energy evolution with respect to time in molecular dynamics simulation. The insets are the top and side views of final configurations for monolayer Nb$_2$SeTeO at 300 K after 5 ps. (d) The band structure without SOC. (e) Orbital projection of the bands without SOC. (f) The spin-resolved band structure and SHC of monolayer Nb$_2$SeTeO with SOC. Inset shows the Berry curvature component along the $z$ direction of magnetization. (g) The edge states of monolayer Nb$_2$SeTeO cut along the [100] direction.
		} \label{NSTO}
	\end{center}
\end{figure}

{\it Candidate materials.} The monolayer Nb$_2$SeTeO takes a square lattice structure with space group of P4mm (No. 99), which is affiliated to the C$_{4v}$ point group symmetry, as shown in Fig. \ref{NSTO}(a). It is a type of Janus material with $\mathcal{I}$ symmetry naturally breaking. Our first-principles electronic structure calculations reveal that monolayer Nb$_2$SeTeO has a checkboard antiferromagnetic ground state with optimized lattice constants \mbox{$a$ = $b$ = 4.20 \AA} \cite{Giannozzi2009, Giannozzi2017, Cococcioni2005, Liechtenstein1995} (The calculation results of the magnetic structure are presented in the supplementary materials (SM).). The Nb-Nb square lattice can be divided into two square sublattices, in each of which the magnetic moments of Nb atoms ($\sim$ 1.0 $\mu_B$) take their own ferromagnetic order. The dynamical and thermal stability of monolayer Nb$_2$SeTeO have been checked by the calculations of phonon spectrum and \textit{ab initio} molecular dynamic simulations \cite{Giannozzi2009, Giannozzi2017, DFPT}. The absence of imaginary frequency modes in the phonon spectrum, and well maintained crystal structure and steadily potential energy evolution curve up to the temperature of 300 K all indicate both good dynamical and thermal stability of monolayer Nb$_2$SeTeO, as shown in Figs. \ref{NSTO}(b-c).

Since the magnetic unit cell and crystal primitive cell are the same, monolayer Nb$_2$SeTeO does not possess $\left\{ C_2^{\bot}||\tau \right\}$ spin symmetry. Moreover, due to lacking inversion symmetry, it does not possess $\left\{ C_2^{\bot}||\mathcal{I} \right\}$ spin symmetry either. However, the two Nb atoms with opposite magnetic moments can be connected by $C_4$ symmetry, which implies the existence of $\left\{ C_2^{\bot}|| C_4 \right\}$ spin symmetry. Therefore, monolayer Nb$_2$SeTeO is a $d$-wave altermagnetic material.


On the other hand, as shown in Fig. \ref{NSTO}(d), the Nb atoms in the sublattice with spin-up (spin-down) are connected through O (Se/Te) atoms along x-axis and vice versa along y-axis. Thus, there is the significant difference between the next-nearest neighbor hoppings of Nb atoms respectively with spin up and spin down, which can lead to a large spin splitting in monolayer Nb$_2$SeTeO without SOC. As shown in Fig. \ref{NSTO}(d), our calculations demonstrate that a large spin splitting is indeed present along the $\Gamma$-X and $\Gamma$-Y directions, but the bands along the $\Gamma$-M direction are spin-degenerate due to $\left\{ C_2^{\bot}|| M_{xy} \right\}$ spin symmetry, which reflects the characteristics of $d$-wave altermagnetism. More importantly, at the Fermi level, monolayer Nb$_2$SeTeO has only two pairs of Weyl points with opposite spins on the X-M and Y-M axes, which are protected by $\left\{ E||M_{x} \right\}$ and $\left\{ E||M_{y} \right\}$ spin symmetries. Therefore, monolayer Nb$_2$SeTeO is a perfect bipolarized Weyl semimetal, which provides a possibility for the realization of type-II QSHE in altermagnetic materials. Meanwhile, the orbital-weight analysis (Fig. \ref{NSTO}(e)) indicates that the bands near the Weyl points are mainly contributed by the $t_{2g}$ orbitals of Nb, which implies that monolayer Nb$_2$SeTeO has a non-negligible SOC effect near the Fermi level.

When SOC is taken into account, the symmetry of monolayer Nb$_2$SeTeO will change from the spin group to the magnetic point group. The specific magnetic point group symmetry depends on the easy magnetization direction of monolayer Nb$_2$SeTeO. Our calculations show that the easy magnetization axis is along the [001] direction. Thus, the symmetry of monolayer Nb$_2$SeTeO is $E$, 2$C_{4z}\mathcal{T}$, $C_{2z}$, $M_x\mathcal{T}$, $M_y\mathcal{T}$, and 2$M_{xy}$. Obviously, the change of symmetry from $\left\{ E||M_{x} \right\}$ ($\left\{ E||M_{y} \right\}$) to $M_x\mathcal{T}$ ($M_y\mathcal{T}$) leads to the opening of band gap at the Weyl point due to band repulsion, as shown in Fig. \ref{NSTO}(f). In a 2D system, a massive Weyl point contributes $\pi$ Berry phase. Due to the $C_{2z}$ symmetry, a pair of Weyl points on the X-M (Y-M) axis contribute the same Berry phase. The $C_{4z}\mathcal{T}$ ($M_{xy}$) symmetry causes the two pairs of Weyl points respectively on the X-M and Y-M axes to contribute opposite Berry phases (Fig. \ref{NSTO}(f)). To be specific, the pair of Weyl points on the X-M axis contributes Chern number being 1, while the pair of Weyl points on the Y-M axis contributes Chern number being -1. Moreover, the spin polarization on the X-M and Y-M axes remains opposite (Fig. \ref{NSTO}(f)), that is, the bands with opposite spin polarization contribute opposite Chern numbers. Then, we also calculated the topological edge states of monolayer Nb$_2$SeTeO, as shown in Fig. \ref{NSTO}(g). The edge states with opposite chirality are located near the $\tilde{\Gamma}$ point and the $\tilde{X}$ point, respectively, which is consistent with Fig. \ref{QSHI}(b). Thus, monolayer Nb$_2$SeTeO is a type-II QSHI.

\begin{figure}
	\centering
		\includegraphics[width=1.0\columnwidth]{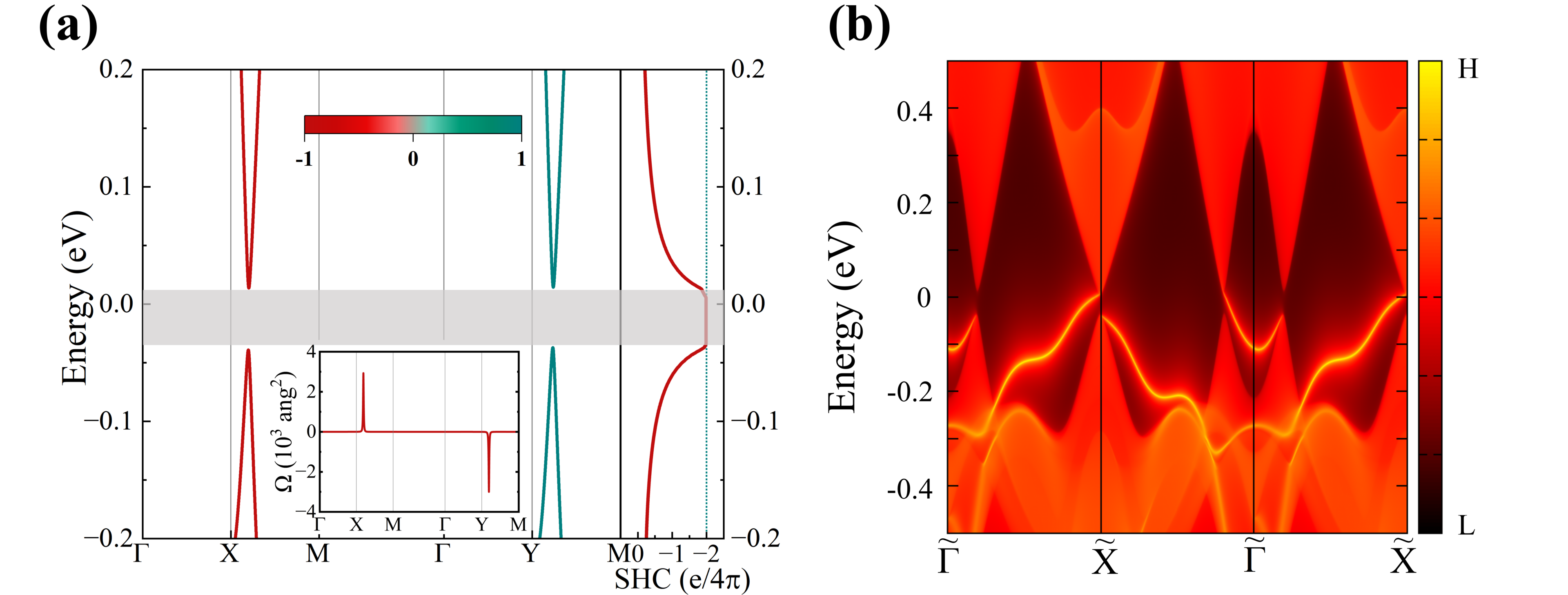}
		\caption{For monolayer Nb$_2$SeTeO under shear strain without any symmetry except translation: (a) The spin-resolved band structure and SHC with SOC. Inset shows the Berry curvature component along the $z$ direction of magnetization. (b) The edge states cut along the [100] direction. 
		} \label{NSTO-distorted}
\end{figure} 

According to the above analysis, the opposite Chern numbers of the spin-polarized bands are jointly guaranteed by the $C_{2z}$ and $C_4\mathcal{T}$ ($M_{xy}$) symmetries. If these symmetries are all broken, can type-II QSHI phase still be realized? Through shear strain, all symmetries of monolayer Nb$_2$SeTeO are broken other than translational symmetry, but its total magnetic momentum is still zero. Thus, the monolayer Nb$_2$SeTeO under sheer stain is a Luttinger compensated magnetic material\cite{PRX-LCM, Liu-PRL, Guo-LCM}. Interestingly, when SOC is considered, the two pairs of Weyl points of monolayer Nb$_2$SeTeO under sheer strain still contribute opposite Chern numbers, as shown in Fig. \ref{NSTO-distorted}(a). Correspondingly, monolayer Nb$_2$SeTeO also has two topological edge states with opposite chirality, which are located near the $\tilde{\Gamma}$ point and the $\tilde{X}$ point, respectively, as shown in Fig. \ref{NSTO-distorted}(b). Meanwhile, our calculations give that the spin Hall conductance (SHC) is quantized to $|\sigma^{z}_{xy}| = 2 e/4\pi$ for monolayer Nb$_2$SeTeO both with (Fig. \ref{NSTO}(f)) and without (Fig. \ref{NSTO-distorted}(a)) shear strain. Therefore, type-II QSHI phase does not require any symmetry protection other than translational symmetry and can be realized in altermagnetic materials and Luttinger compensated magnetic materials.

In addition, the realization of the QSHE requires that the X-M and Y-M directions still have opposite spin polarization. If SOC is very strong, will the QSHI still be stable? Next, we will use a lattice model to address the question.

\begin{figure*}[htp]
	\centering
	\includegraphics[width=16cm]{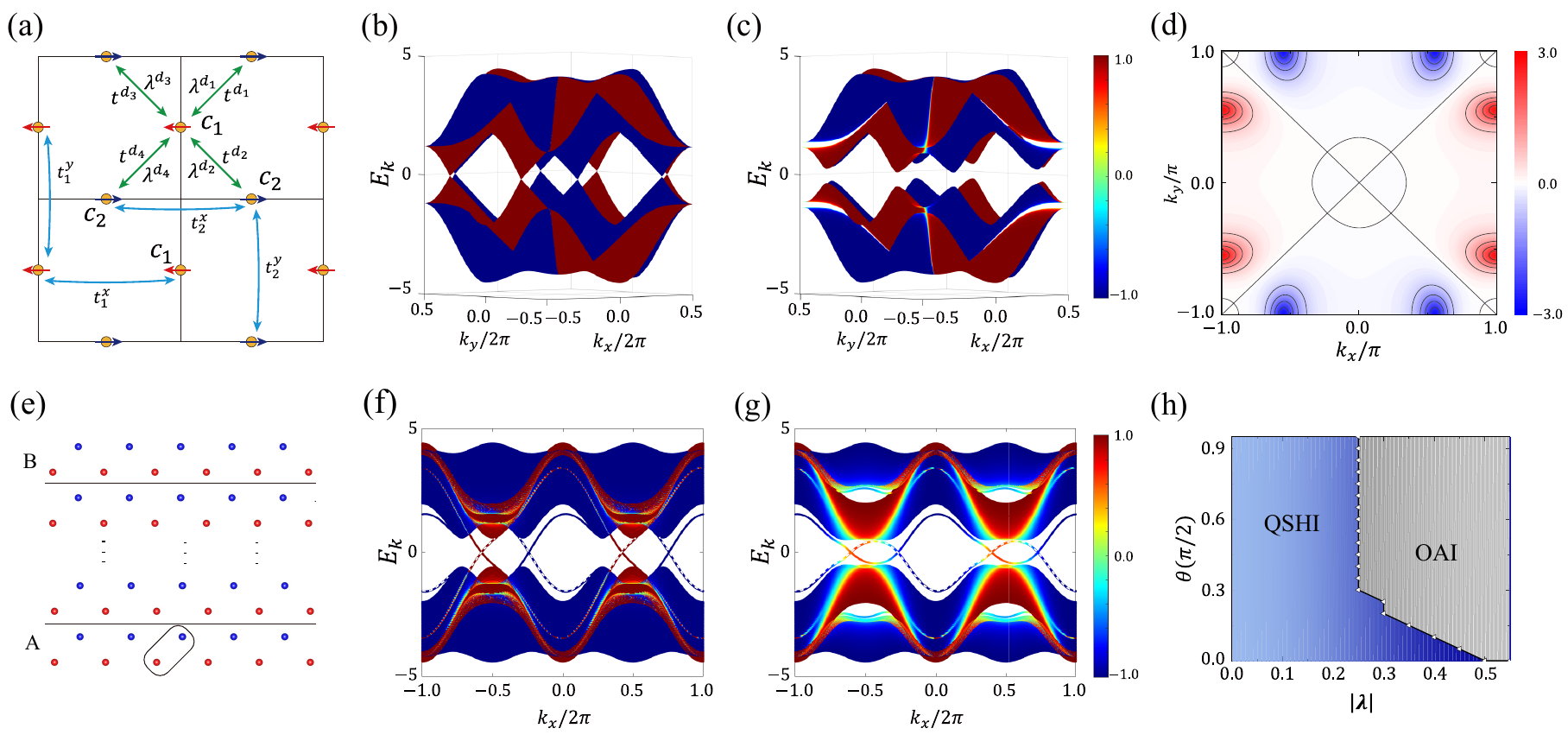}
	\caption{The 2D lattice model with altermagnetism. (a) The schematic illustration for the hoppings between lattice sites for the Hamiltonian Eq. (\ref{Eq1}). The red and blue arrows represent spin-up and spin-down magentic moments, respectively. The band structure of lattice model without SOC (b) and with SOC for $\lambda=0.2$, $\theta=0.2\times\pi/2$ (c). (d) The Berry curvature for the band structure in (c). (e) represents the case with open boundary conditions along the $y$ direction while maintaining periodic boundary conditions in the $x$ direction. (f) are the corresponding edge states with the projection of $s_z$ for (c).  The edge states for obstructed atomic insulator state (h) with $(|\boldsymbol{\lambda}|, \theta)=(0.4,0.8\times\pi/2)$. Other parameters: $t^d=1$, $t_1=-t_2=0.8$, $|\mathbf{m}|=m_z=1.2$. The solid lines and dashed lines represent the edge states of the A cut and B cut boundaries (The detailed results are presented in the SM), respectively. (h) The phase diagram of lattice model.}
	\label{fig4}
\end{figure*}


{\it Lattice model.} In order to address the above question, we consider a tight-binding (TB) model, in which a square lattice contains two sites (labeled by sublattice index $\alpha=1, 2$) within one unit cell. The corresponding model Hamiltonian is \cite{tan2024bipolarizedweylsemimetalsquantum}
\begin{align} 
H = &  \sum_{d_i,j} \left[ t^{d} C_{1,j}^\dagger C_{2,j+\bm{d}_i} + C_{1,j}^\dagger (i\boldsymbol{\lambda^d}\cdot \boldsymbol{\sigma}) C_{2,j+\bm{d}_i} + h.c.\right] \notag\\
    &+ \sum_{\alpha,j} \left[ t^{x}_{\alpha} C_{\alpha,j}^\dagger C_{\alpha,j+\mathbf{x}} + t^{y}_{\alpha} C_{\alpha,j}^\dagger C_{\alpha,j+\mathbf{y}} + h.c.\right] \notag\\
    &+ \sum_{\alpha,j}  \left.\bf{m}_\alpha \cdot \boldsymbol{\sigma} \right. C_{\alpha,j}^\dagger C_{\alpha,j},
\label{Eq1}
\end{align}
where $ C_{\alpha,j}^\dagger = \left( C_{\alpha,j\uparrow}^\dagger, C_{\alpha,j\downarrow}^\dagger \right) $ and $ C_{\alpha,j} = \left( C_{\alpha,j\uparrow}, C_{\alpha,j\downarrow} \right) $ represent electron creation and annihilation operators, respectively. As illustrated in Fig.~\ref{fig4}(a), $t^{d}$ and $\bm{d}_{1,4} = \pm\frac{1}{2}(\mathbf{x}+\mathbf{y}), \bm{d}_{2,3} = \pm\frac{1}{2}(\mathbf{x}-\mathbf{y})$, $t^{x,y}_\alpha$ and $\mathbf{x} = a_1 \hat{x}, \mathbf{y} = a_2 \hat{y}$ represent the strength and direction of the nearest and next nearest hopping, respectively, where $a_1, a_2$ are the lattice constants along the unit vectors $\hat{x}, \hat{y}$. For the SOC term, we consider $\lambda^{d_1}=\lambda^{d_3}$, $\boldsymbol{\lambda}^{d_2}=\boldsymbol{\lambda}^{d_4}$, and $\boldsymbol{\lambda}^{d_1}=-\boldsymbol{\lambda}^{d_2}=\boldsymbol{\lambda}$. The $\sigma_0$ and $\boldsymbol{\sigma}$ are identity matrix and Pauli matrix, respectively, while $\textbf{m}_1=-\textbf{m}_2=\textbf{m}$ stands for the static checkboard antiferromagnetic order which couples to the electron spin as a Zeeman field. Further, $t_1^{x,y}\neq t_2^{x,y}$ and $t_1^{x,y} = t_2^{y,x}$ break the spin symmetry $\left\{C^{\bot}_2||\tau\right\}$ but preserves spin symmetry $\left\{C^{\bot}_2||C_{4z}\right\}$ and $\left\{C^{\bot}_2||M_{xy}\right\}$. Since the magnetic lattice points are located at inversion-symmetric points, this lattice model also lacks spin symmetry $\left\{C^{\bot}_2||\mathcal{I}\right\}$. Thus, this lattice model is a $d$-wave altermagnetism. 

After performing the Fourier transformation, the model Hamiltonian can be rewritten as $H=\sum_k\psi_k^\dagger H_0 \psi_k$ in momentum space with basis $\psi^\dagger = \left( C_{1k\uparrow}^\dagger, C_{1k\downarrow}^\dagger, C_{2k\uparrow}^\dagger, C_{2k\downarrow}^\dagger \right)$, where $H_0$ reads
\begin{align}
	H_0 =\Gamma^{+}_k \tau_0 \sigma_0 + \Gamma_k^{12} \tau_x \sigma_0+\Gamma_k^{-} \tau_z \sigma_0 + \Gamma_s \tau_y \boldsymbol{\lambda}\cdot\boldsymbol{\sigma} + \tau_z \bf{m} \cdotp \boldsymbol{\sigma}
 \label{Eq2}
\end{align}
in which Pauli matrix $\tau_0$ and $\boldsymbol{\tau}$ are identity matrix and pseudospin matrix in lattice space, respectively. The auxiliary functions $\Gamma_k^{12} = 4t^d \cos\frac{k_x}{2} \cos\frac{k_y}{2}$, $\Gamma_s = 4 \sin \frac{kx}{2} \sin \frac{ky}{2}$ and $\Gamma_k^{\pm}=(t_1 \pm t_2)\cos k_x + (t_2 \pm t_1)\cos k_y$ with $t_\alpha^x=t_\alpha$  are related to inter- and intra- sublattice hopping, respectively. The eigenvalues of Eq. (\ref{Eq2}) read  
\begin{align}
\varepsilon_{\pm}^{\pm}(\boldsymbol{k}) = \Gamma^{+} \pm \sqrt{(\Gamma_k^{12})^2 + \left| \mathbf{m} \right|^2 + (\Gamma^{-})^2 + \Gamma_s^2 \left| \boldsymbol{\lambda} \right|^2 \pm 2 A}.
 \label{Eq3}
\end{align}
with $A= \sqrt{(\left| \mathbf{m} \right| \Gamma^{-})^2+\Gamma_s^2\left|| \mathbf{m} \times \boldsymbol{\lambda} \right||^2 }$.

For $\boldsymbol{\lambda}=0$, namely without SOC, the TB model (Eq. \ref{Eq1}) is type-I bipolarized Weyl semimetals with two pairs of opposite spin polarized Weyl cones, which are protected by the spin symmetry $\left\{ E||C_{2y} \right\}$ and $\left\{ E||C_{2x} \right\}$, respectively, as shown in Fig. \ref{fig4}(b) \cite{tan2024bipolarizedweylsemimetalsquantum}. Therefore, the electronic structure of this model is consistent with that of monolayer Nb$_2$SeTeO in the absence of SOC. To better investigate the properties of the model under SOC, we define the angle between $\mathbf{m}$ and $\boldsymbol{\lambda}$ as $\theta$, satisfying $\cos \theta=\frac{\mathbf{m}\cdot\boldsymbol{\lambda}}{|\mathbf{m}||\boldsymbol{\lambda}|}$ with $\theta\in [0,\frac{\pi}{2})$.

If $\mathbf{m} \parallel \boldsymbol{\lambda}$ ($\theta=0$), all Weyl cones open the gaps due to band repulsion, resulting in type-II QSHI. Due to the presence of $U(1)$ symmetry, spin $S_z$ remains a good quantum number. Thus, type-II QSHI phase is always stable with the enhancement of SOC, as shown in Fig. \ref{fig4}(h). 
Once $\mathbf{m} \nparallel \boldsymbol{\lambda}$ ($0<\theta<\pi/2$), the $U(1)$ symmetry is explicitly broken and spin is no longer a good quantum number. With the enhancement of SOC, this model transitions from type-II QSHI phase to OAI phase, as demonstrated by our calculations of Berry curvature (Fig. \ref{fig4}(d)) and the edge states. In the blue region of Fig. \ref{fig4}(h), the edge states with opposite chirality connect the valence and conduction bands (Fig. \ref{fig4}(f)), corresponding to type-II QSHI phase. In the gray region of Fig. \ref{fig4}(h), the edge states lie completely within the band gap (Fig. \ref{fig4}(g)), corresponding to the OAI phase. Since no symmetry is broken during the phase transition, the phase transition is a topological phase transition. On the other hand, although realistic materials always break the $U(1)$ symmetry, our calculations show that the type-II QSHI phase can emerge within a relatively broad range of parameters. Therefore, type-II QSHI phase can be realized in altermagnetic and Luttinger compensated magnetic materials, for example monolayer Nb$_2$SeTeO.


In summary, based on theoretical analysis, we propose type-II QSHI, which has edge states with opposite chirality and polarization distributed in different regions of the Brillouin zone and quantized spin Hall conductivity. Subsequently, through symmetry analysis and the first-principles electronic structure calculations, we demonstrate that the type-II QSHI can exist not only in altermagnetic materials but also in Luttinger compensated magnetic materials. Furthermore, based on lattice model calculations, we show that even with the breaking of $U(1)$ symmetry, type-II QSHI can still exist over a broad range of parameters. With the enhancement of SOC, type-II QSHI phase will transition to the OAI phase, and the phase transition is a topological phase transition. Our work proposes a new mechanism for realizing the QSHE and also enriches the topological classes of unconventional magnetism.


This work was financially supported by the National Natural Science Foundation of China (Grants Nos. 12074040, 12204533, 12434009, 62206299, 12174443, 12274255, 11974207, and 11974194), the National Key R$\&$D Program of China (Grant No.2024YFA1408601), the Fundamental Research Funds for the Central Universities, and the Research Funds of Renmin University of China (Grant No. 24XNKJ15), the Beijing Natural Science Foundation (Grant No.Z200005) and the Innovation Program for Quantum Science and Technology (2021ZD0302402). F. Ma was also supported by the BNU Tang Scholar. The computations were supported by the Center for Advanced Quantum Studies at Beijing Normal University, and the Physical Laboratory of High Performance Computing at Renmin University of China.

\bibliography{reference}

\end{document}